\documentclass[twocolumn,prb,showpacs,floatfix,superscriptaddress]{revtex4}
\usepackage{graphicx}
\begin{document}
\title{Quantum Monte Carlo simulation of  thin magnetic films}

\author{P. Henelius} 
\email{patrik@theophys.kth.se}
\affiliation{Condensed Matter Theory, Royal Institute of Technology,
SE-106 91 Stockholm, Sweden}

\author{P. Fr{\"o}brich} 
\affiliation{Hahn-Meitner-Institut Berlin, Glienicker Stra{\ss}e 100,
14109 Berlin, Germany}
\affiliation{Institut f{\"u}r Theoretische Physik, Freie
Universit{\"a}t Berlin, Arnimallee 14, 14195 Berlin, Germany}

\author{P.J. Kuntz} 
\affiliation{Hahn-Meitner-Institut Berlin, Glienicker Stra{\ss}e
100,
 14109 Berlin, Germany}

\author{C. Timm}
\affiliation{Institut f{\"u}r Theoretische Physik, Freie
Universit{\"a}t Berlin, Arnimallee 14, 14195 Berlin, Germany}

\author{P.J. Jensen}
\affiliation{Institut f{\"u}r Theoretische Physik, Freie
Universit{\"a}t Berlin, Arnimallee 14, 14195 Berlin, Germany}
\affiliation{Laboratoire de Physique Quantique, Universite Paul
Sabatier, UMR 5626 du CNRS 118, route de Narbonne, F-31062 Toulouse, France}

\date{\today}

\begin{abstract}
The stochastic series expansion quantum Monte Carlo method is used to
study thin ferromagnetic films, described by a Heisenberg model
including local anisotropies. The magnetization curve is calculated,
and the results compared to Schwinger boson and many-body Green's
function calculations. A transverse field is introduced in order to
study the reorientation effect, in which the magnetization changes
from out-of-plane to in-plane. Since the approximate theoretical
approaches above differ significantly from each other, and the Monte
Carlo method is free of systematic errors, the calculation provides an
unbiased check of the approximate treatments. By studying quantum spin
models with local anisotropies, varying spin size, and a transverse
field, we also demonstrate the general applicability of the recent
cluster-loop formulation of the stochastic series expansion quantum
Monte Carlo method.
\end{abstract}

\maketitle

\section{Introduction}

The driving force behind much of the research on thin magnetic films
is their application in data storage devices. Magnetic thin films also
display many remarkable physical phenomena, such as the reorientation
effect,\cite{blan94} in which the axis of magnetization changes as a
function of film thickness, temperature and applied fields.  Many
theoretical methods, such as ab-initio calculations,\cite{hjor97}
mean-field theories,\cite{mosc94, jb98} classical Monte Carlo
simulations,\cite{eric91, igle99} Green's function
methods\cite{jens99, eck99, frob01,frob00,frob02,eric91b} and
Schwinger bosons\cite{timm00} have been applied to thin-film
systems. Much of this work has focused on using the Heisenberg model
to study the reorientation effect. The ground state and the lowest
(one-magnon) excitations  are known for the two-dimensional
ferromagnetic Heisenberg model, but there is no closed form analytic
solution at finite temperatures. Since results obtained using
different approximate methods differ a great deal from each other,
there is a need for an unbiased check of the various methods used;
this is achieved by quantum Monte Carlo (QMC) calculations. In
Ref.~\onlinecite{eck99}, for example, it has been shown by comparing
with QMC results\cite{tghs98,hene00a} that the Tyablicov \cite{tyab59}
decoupling (random phase approximation: RPA) is a very good
approximation for
 the magnetization of a spin $S=1/2$ monolayer in
an external magnetic field (perpendicular to the film plane). The main
purpose of the present paper is to show the feasibility of large-scale
quantum Monte Carlo (QMC) calculations, free of systematic errors, in
the study of thin magnetic films. In order to achieve this goal we
have included higher spins, local anisotropies, and a transverse
magnetic field in the operator-loop formulation of the stochastic
series expansion (SSE) QMC method.\cite{sand99}

Two often-cited trends in quantum Monte Carlo development are the
emergence of methods free of systematic errors,\cite{sand91, prok96,
bear96, romb99, hene00a} and the development of highly efficient
loop-cluster algorithms.\cite{ever93, kawa94, prok96, bear96, sand99,
sylj02} In this work we particularly want to emphasize the
\textit{general applicability} of the SSE operator-loop method, which
makes it possible to use the same algorithm to study a wide variety of
Hamiltonians.  Whereas previously it was necessary to rewrite large
sections of the computer code when changing the model, one can now use
the same code (compiled only once) to simulate a wide range of
different systems. The user of the program no longer necessarily needs
detailed knowledge of the algorithm and code to be able to conduct a
thorough study of many quantum spin models in any dimension with, for
example, non-zero magnetic fields, anisotropies and varying spin size.

A brief description of the QMC method is given in Sec.~\ref{sec:sse}.
Thereafter we discuss the applicability of the method and the
introduction of general spin size and a transverse field in the SSE
operator-loop algorithm.  In Sec.~\ref{sec:res} we compare some
examples of the QMC simulations to approximate theoretical
approaches. Finally, we comment on possible future applications of QMC
in the context of thin magnetic films.

\section{\label{sec:sse}SSE cluster-loop algorithm}

There are excellent descriptions of the SSE loop
algorithm,\cite{sand99,sylj02} so we only give a brief summary here in
order to introduce the general framework of the method. We will,
however, try to describe the main features of the method
pictorially. The focus is on the new aspects that arise when
introducing arbitrary spin size and a transverse field.

We consider a lattice spin model described by a Hamiltonian $H$.  The
SSE method relies on a Taylor expansion of the partition function $Z$:
\begin{equation}
Z=\sum_{\alpha}\sum_{n=0}^{\infty} \frac{(-\beta)^n}{n!}\langle\alpha
| H^n |\alpha\rangle,
\end{equation}
where $|\alpha\rangle$ are basis states in which the matrix element
above can be evaluated, and $\beta$ is the inverse temperature.

To describe the updating procedure, we write the Hamiltonian as a sum
over all $M$ bonds representing interacting spins in the system
\begin{equation}
H=-\sum_{b=1}^M H_b.
\end{equation}
The bond operator $H_b$ can be decomposed into its diagonal and
off-diagonal parts,
\begin{equation}
	H_b=H_{D,b}+H_{O,b},
\end{equation}
where subscript $D$ denotes a diagonal operator and $O$ an
off-diagonal operator. For a ferromagnetic Heisenberg model these two
operators are of the form
\begin{equation}
-H_{D,b}=S_{i(b)}^zS_{j(b)}^z
\end{equation}
and
\begin{equation}
-H_{O,b}=S_{i(b)}^+S_{j(b)}^-+S_{i(b)}^-S_{j(b)}^+,
\end{equation}
where $i(b)$ and $j(b)$ denote the two spins connected by bond $b$.
If we introduce a cutoff order $L$ in the Taylor expansion (which,
when done properly, does not cause any systematic
errors\cite{sylj02}), and include additional unit operators $I$, the
expansion can be rewritten in the form
\begin{equation}
H=\sum_{\alpha}\sum_{S_L} \frac{\beta^n(L-n)!}{L!}\langle\alpha
| S_L |\alpha\rangle,
\label{eq:taylor}
\end{equation}
where $S_L$ is an operator string
\begin{equation}
S_L=\prod_{p=1}^L H_p,
\end{equation}
with $H_p\in \{H_{D,b}, H_{O,b}, I\}$.  The Monte Carlo procedure must
sample the space of all states $|\alpha\rangle$ and all operator
sequences $S_L$.

\begin{figure}
\resizebox{!}{4cm}{\includegraphics{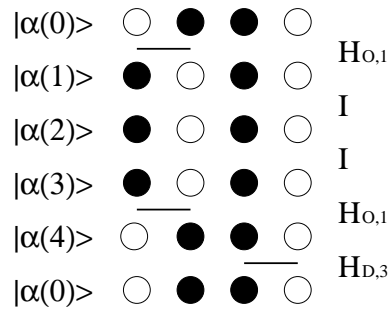}}
\caption{\label{fig:conf} An SSE configuration for a four-site
spin-1/2 model. A filled (empty) circle denotes a state with spin up
(down). A horizontal bar indicates a vertex, corresponding to a bond
operator $H_b$, labeled on the right. The propagated states
$|\alpha(p)\rangle$ are labeled on the left side.}
\end{figure}

We next consider the SSE space in more detail. Denoting a propagated
state by
\begin{equation}
|\alpha(p)\rangle=\prod_{i=1}^pH_i|\alpha\rangle,
\end{equation}
the matrix element in Eq.~(\ref{eq:taylor}) can be written as a
product of elements of the form
$\langle\alpha(p)|H_b|\alpha(p-1)\rangle$, where the bond-operator
$H_b$ only acts on two spins.  For the spin-1/2 Heisenberg model the
only elements that can appear are $\langle\uparrow
,\uparrow|H_b|\uparrow ,\uparrow\rangle$, $\langle\uparrow
,\downarrow|H_b|\uparrow ,\downarrow\rangle$, $\langle\uparrow
,\downarrow|H_b|\downarrow ,\uparrow\rangle$, and the spin-reversed
versions of the same set. From now on we refer to these matrix
elements, consisting of four spin states and a bond operator, as
``vertices''. The matrix element in Eq.~(\ref{eq:taylor}) can thus be
viewed as a list of vertices of the above kind.  In
Fig.~\ref{fig:conf} we depict such a list graphically.

\begin{figure}
\resizebox{!}{4cm}{\includegraphics{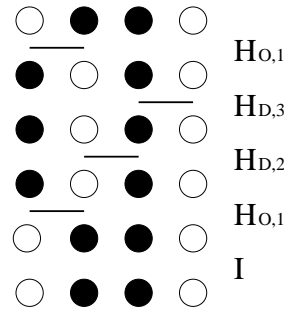}}
\caption{\label{fig:dia}
A possible outcome of performing a diagonal update on the configuration
in Fig.~\ref{fig:conf}.}
\end{figure}

In the operator-loop algorithm, two basic updates ensure that the
complete SSE space is sampled. The diagonal update attempts to
exchange diagonal operators $H_{D,b}$ and unit operators $I$.  The
probability for inserting a diagonal operator (exchanging it for a
unit operator) at position $p$ in the operator sequence is
\begin{equation}
P_{\text{insert}}=\frac{M\beta \langle\alpha(p)|
H_{D,b}|\alpha(p)\rangle}{L-n},
\label{eq:ins}
\end{equation}
while the probability for removing a diagonal operator is
\begin{equation}
P_{\text{remove}}=\frac{L-n+1}{M\beta \langle\alpha(p)|
H_{D,b}|\alpha(p)\rangle}.
\label{eq:rem}
\end{equation}
In a diagonal update, one exchange attempt is made for each diagonal
and unit operator. A typical outcome of a diagonal update is shown in
Fig.~\ref{fig:dia}.

The second type of update is a global operator-loop update, which
leaves unit operators unaffected. The idea of the loop move is to form
and flip a closed loop of spins in the vertex list. In the process
both the affected vertices and states are changed. The operator-loop
update together with the above diagonal update ensure that the
complete SSE configuration space is sampled.

\begin{figure}
\resizebox{!}{8cm}{\includegraphics{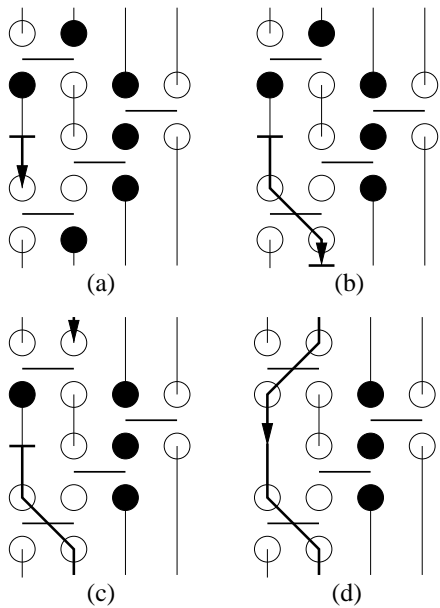}}
\caption{\label{fig:loop} A loop update. Unit operators have been
removed from the configuration in Fig.~\ref{fig:dia}. Vertical
lines (links) show how the spins at different vertices are
connected. Short horizontal bars denote link-discontinuities.}
\end{figure}

The loop move is easy to formulate. In Fig.~\ref{fig:loop} we
illustrate a loop move graphically.  We have removed all the unit
operators $I$ from the operator sequence in Fig.~\ref{fig:dia}.
Furthermore we only show the spins that are members of a vertex, and
we turn the configuration into a linked list by connecting the same
spin at different vertices by vertical lines, which we refer to as
``links''. Next we describe the loop move in more detail.

First a random spin, belonging to a vertex, is selected and flipped,
see Fig.~\ref{fig:loop}(a). We refer to this spin as the ``entrance''
spin to the first vertex. For the loop to proceed, we select an
``exit'' spin, also belonging to the same vertex, see
Fig.~\ref{fig:loop}(b).  The probability of choosing a given exit spin
is proportional to the vertex matrix element, that results from
flipping the entrance and exit legs. This choice of exit spin ensures
detailed balance.\cite{sylj02}The exit spin is linked to another spin
in the linked vertex list, and this spin is chosen as the entrance
spin to the next vertex, see Fig.~\ref{fig:loop}(c). In this manner a
loop of spins on the space-time lattice is formed and flipped.  Note
that when the first spin is flipped a discontinuity is introduced into
the vertex list.  A link-discontinuity appears when two spins linked
together are not in the same state. When, for example, the initial
entrance and exit spins (assuming they are different) are flipped, two
discontinuities are introduced in the linked list, see
Fig.~\ref{fig:loop}(b), where a link-discontinuity is marked by a
short horizontal bar. One of the discontinuities gets propagated by
the loop until the loop passes through the initial spin a second time,
when the discontinuities ``annihilate'' each other and the loop
closes, see Fig.~\ref{fig:loop}(d). A practical and sufficient
criterion for closing the loop is that the exit spin (after having
been flipped) is in the same state as the next entrance spin (before
flipping it). When this is the case, all discontinuities have been
removed and the loop can be closed.

\subsection{Applicability of the algorithm}

The SSE loop algorithm is extremely general and can be efficiently
applied to any model with a positive definite partition function (if
this is not the case, one encounters the usual difficulties due to the
sign problem\cite{hene00b}). This means that the algorithm can be
applied to ferromagnetic and antiferromagnetic (on bipartite lattices)
Heisenberg models of any quantum spin size in any dimension, including
magnetic fields and local as well as exchange anisotropies. The
interactions need not be short-ranged, as long as they do not cause
frustration (which leads to the sign problem).

From a programming point of view, one need only calculate the energies
of all possible vertices (which typically is a simple task) and
generate a table of exit probabilities as a function of four
variables: the initial vertex type, the entrance spin, the entrance
spin state and the exit spin.

As an illustration we have written a code that works for a general
$d$-dimensional Heisenberg model of the form
\begin{eqnarray}
H=&&-\sum_{ij} \left[J_{ij}^zS_i^zS_j^z+
J_{ij}^t(S_i^+S_j^-+S_i^-S_j^+)\right]\nonumber\\
&& -\sum_i \left[K_2 (S_i^z)^2
+K_4 (S_i^z)^4\right]\nonumber\\
&& -\sum_i \left[B^zS_i^z+B^xS_i^x\right],
\label{eq:model}
\end{eqnarray}
 where the exchange interaction is given by $J_{ij}$ (negative for
antiferromagnets and positive for ferromagnets). $\mathbf{S}$ denotes the
quantum spin and $K_2$ and $K_4$ are the second and fourth-order local
uniaxial anisotropies. $\mathbf{B}$ is an external magnetic field. In
this study we limit ourselves to a nearest neighbour interaction, but
in principle one can include long-range interactions. In the following
we describe the modifications to the loop algorithm that allow the
introduction of general spin size and a transverse field.

\subsection{\label{sec:spin}Inclusion of arbitrary spin}
For the sake of simplicity, the SSE loop-operator algorithm was
originally described\cite{sand99} for a spin-1/2 system, but the
generalization to a spin-$S$ system is straightforward. For an
arbitrary spin system there are more allowed vertices, but the
probability for inserting and extracting diagonal operators is still
given by Eq.~(\ref{eq:ins}) and Eq.~(\ref{eq:rem}) respectively.  The
only necessary change in the loop-update is the choice of the initial
spin state. For the spin-1/2 model the initial entrance spin was
simply flipped, but for the spin-$S$ model one can randomly choose
among the $2S$ spin states that differ from the initial spin
state. All other aspects of the algorithm remain the same.

\subsection{\label{sec:field}Inclusion of a transverse field}
Previously a transverse field has been included in the SSE algorithm
using local updates.\cite{hene00a} In the present work we include the
transverse field directly in the global loop update. There are very
likely many different ways to formulate a loop-update which includes a
transverse magnetic field.  Here we will not make an exhaustive study,
but rather present one possible algorithm, which we found particularly
simple to implement. It is probably not the most efficient update, and
we are currently making a more detailed study of more efficient
algorithms for including the transverse field.

Including a transverse magnetic field is more involved because the
total $z$ component of the magnetization is no longer a conserved
quantity, $[H,\sum_i S_i^z]\neq 0$, due to the presence of single
lowering and raising operators ($2S^x=S^++S^-$) in the Hamiltonian. As
a consequence additional ``flip vertices'', with different
magnetization in the initial and final states (such as
$\langle\uparrow ,\uparrow|H_b|\uparrow ,\downarrow\rangle$), are also
allowed in the vertex-list.  The presence of flip vertices implies
that, in general, one can exit the same exit spin with several
different spin states. The increase in the number of exit states means
that the exit probability is a function of five variables instead of
four: original vertex, entrance spin, entrance state, exit spin and
exit state.  For a model that conserves the total $z$ component of the
magnetization the exit state is a function of the other four
variables.

\begin{figure}
\resizebox{!}{8cm}{\includegraphics{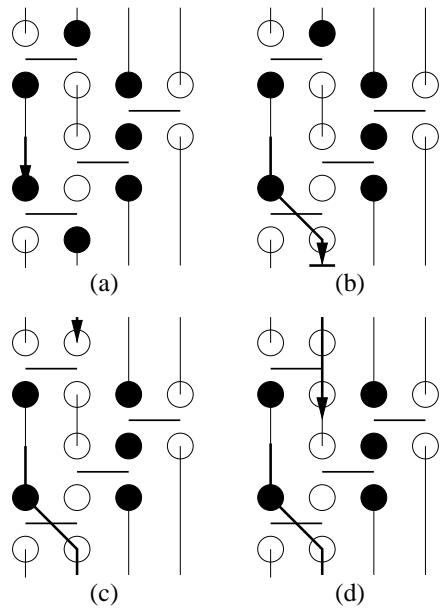}}
\caption{\label{fig:flip} A loop update in the presence of a
transverse field. Note that there is no initial link-discontinuity,
since the state of the initial spin in (a) is left unchanged. With a
transverse field the first and last spins in the loop do not
necessarily coincide.}
\end{figure}

To include a magnetic field, we also have to reconsider how to start,
and end, the operator-loop. With no magnetic field we introduce an
initial link-discontinuity by flipping the first entrance spin. A
second link-discontinuity appears when the first exit spin is flipped
(unless the entrance and exit spins coincide, in which case the loop
closes immediately). The second discontinuity is propagated as the
loop progresses until the two discontinuities annihilate each other
and the loop closes.  With flip vertices allowed, the number of
link-discontinuities can change by the introduction or removal of
single flip vertices.  If one therefore chooses to start, and end, the
loop in the same way as previously described, the loop can close
without having corrected for the initial discontinuity. In such a case
it would be necessary to return to the starting point of the loop and
continue execution until the last discontinuity is removed.

Here we chose a slightly different approach, which can be implemented
by a very minor change in the original formulation. Instead of picking
an initial entrance spin and changing its state we leave its state
\textit{unchanged}.  Without a magnetic field the loop would close
immediately. With a magnetic field this possibility still exists, but
the loop may also start with a flip vertex and thereafter proceed in a
``normal'' fashion until another flip vertex is inserted, or removed,
when the loop is closed according to the usual criterion (the exit
spin and the following entrance spin are in the same state). We have
compared high-precision QMC data for small systems with exact
diagonalization, to ensure that detailed balance is indeed satisfied,
but we do not provide a strict proof here.

Fig.~\ref{fig:flip} depicts an example of a loop move for the case of
a non-zero transverse field. In Fig.~\ref{fig:flip}(a) an initial
entrance spin is selected and left in its original state. An exit spin
is selected and flipped (Fig.~\ref{fig:flip}(b)), resulting in a flip
vertex. The next entrance spin can be seen in Fig.~\ref{fig:flip}(c),
with a corresponding exit spin in Fig.~\ref{fig:flip}(d). Note that
the exit state in Fig.~\ref{fig:flip}(d) is unchanged, resulting in a
second flip vertex, leading to termination of the loop. The loop does
not close on itself in this case, but starts and ends at two different
spins.

The transverse magnetization is particularly easy to calculate within
the SSE formulation, since it simply is equal to the average number of
flip vertices, $N_{\text{flip}}$, in the operator
sequence,\cite{hene00a}
\begin{equation}
M_x=\frac{1}{\beta}\langle N_{\text{flip}}\rangle.
\end{equation}

To conclude this section we note that only two changes need to be made
to add a transverse field.  First of all the initial entrance spin
state should be left unchanged. Secondly, the exit probability is now
also a function of exit state, since there are, in general, several
possible exit states for a given exit spin. Next we apply this
algorithm to a two-dimensional ferromagnetic system.

\section{\label{sec:res}Application to thin magnetic films}

The simplest effective model for a thin magnetic film is a
ferromagnetic Heisenberg monolayer.  However, the Mermin-Wagner
theorem\cite{merm66} tells us that this model cannot have a finite
critical temperature in two dimensions. The continuous symmetry can be
explicitly broken by an anisotropy in order to induce a finite
critical temperature. Typical experimentally observed values of the
local anisotropy in 3$d$ transition metal films are about two orders
of magnitude smaller than the exchange coupling, and one might
therefore expect a very small critical temperature. However, it turns
out that the critical temperature contains a logarithmic dependence on
the anisotropy,\cite{band88,eric91,timm00} so that very small
anisotropies induce a critical temperature of the order of the
coupling constant. In this section we concentrate on results for the
magnetization as a function of temperature. The model is given by
Eq.~(\ref{eq:model}), where the double sum is over all nearest
neighbours on a two-dimensional square lattice. Since the main purpose
of the present paper is to show the feasibility of the quantum Monte
Carlo method, and not to explore the whole parameter range of the
model as defined in Eq.~(\ref{eq:model}), we limit ourselves to some
illustrative examples. From now on we will assume an isotropic
exchange interaction ($J_{ij}^t=J_{ij}^z/2$), zero vertical magnetic
field ($B^z=0$) and only second-order anisotropy ($K_4=0$). First we
discuss how the finite-size effects were treated. Thereafter we
compare the QMC results with approximate theoretical
approaches. Finally, we also show a case where a transverse field
drives the magnetization in the plane.

\subsection{\label{sec:finite}Finite-size effects}

\begin{figure}
\resizebox{!}{6cm}{\includegraphics{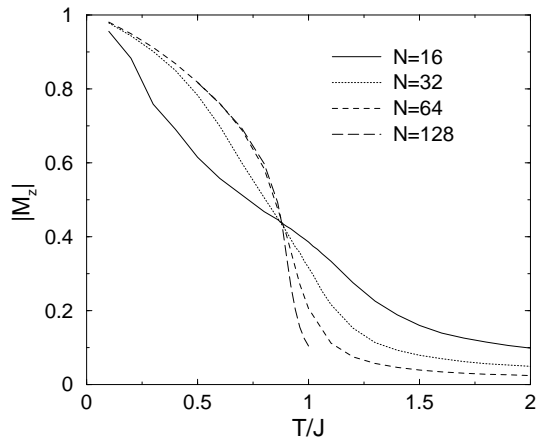}}
\caption{\label{fig:magN} Magnetization as a function of linear system size
$N$ and temperature $T/J$  for $S=1$ and $K_2/J=0.01$.}
\end{figure}

The first problem one encounters in a finite-size system with
spin-inversion symmetry is that the magnetization should vanish
because opposite spin orientations occur equally likely. One can
circumvent this problem by calculating either the absolute value of
the magnetization, or the magnetization squared (and take the root
afterwards).  Both approaches are equivalent in the thermodynamic
limit, and, since the absolute value of the magnetization is found to
converge faster, we show only the former. The magnetization curves for
a particular value of the anisotropy, $K_2/J=0.01$, and different
system sizes are shown in Fig.~\ref{fig:magN}. The finite-size effects
increase closer to the Curie temperature $T_c$, as expected for a
second-order phase transition. For low temperatures, however, one can
see that it is possible to extract data which have converged (within
statistical error) with respect to system size. In the next section we
restrict ourselves to results for the magnetization that show no
discernible finite-size effects.

\begin{figure}
\resizebox{!}{6cm}{\includegraphics{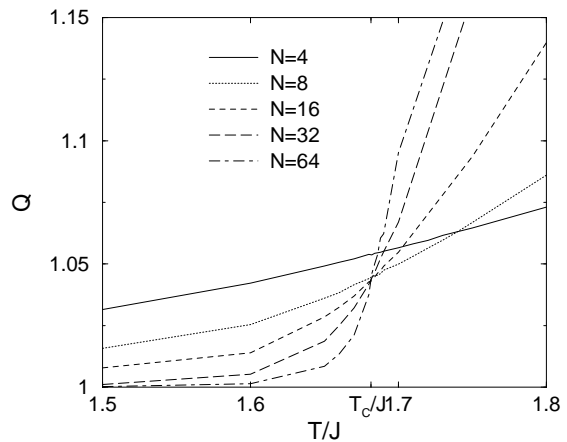}}
\caption{\label{fig:bind150} The Binder ratio $Q$ as a function of linear
system size $N$ and temperature $T/J$ for $S=1$, and $K_2/J=1.5$.}
\end{figure}

The critical temperature can be determined directly. We have used the
Binder ratio\cite{bind81} to extract the value of $T_c$. By plotting
the ratio of two moments of the magnetization,
\begin{equation}
Q=\frac{\langle M_z^4\rangle^\frac{1}{4}}{\langle M_z^2\rangle^\frac{1}{2}},
\end{equation}
the finite-size effects around $T_c$ should largely cancel, and curves
for different system sizes are expected to intersect at $T_c$. In
Fig.~\ref{fig:bind150} we show an example for $K_2/J=1.5$, where one
can clearly see how data for different system sizes intersect at one
point ($T_c/J$).

\begin{figure}
\resizebox{!}{6cm}{\includegraphics{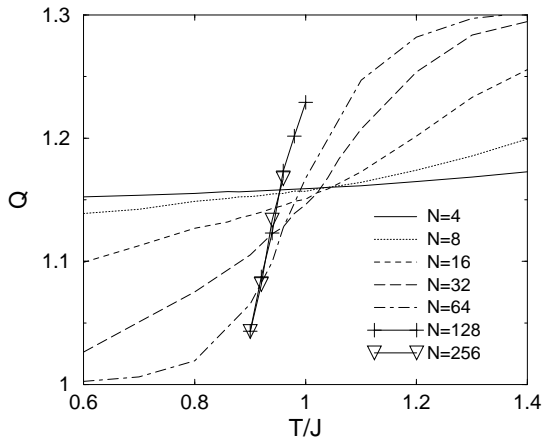}}
\caption{\label{fig:bind001} The Binder ratio $Q$ as a function of linear
system size $N$ temperature $T/J$ for $S=1$ and $K_2/J=0.01$.}
\end{figure}

In transition-metal thin films, the second-order anisotropy is
believed to be of the order $K_2/J=0.01$. The relatively small energy
scale of the anisotropy induces large finite-size effects. In order to
see the effects of a small energy scale one needs to examine large
systems at low temperatures. In Fig.~\ref{fig:bind001} we show the
Binder ratio as a function of temperature and system size for
$K_2/J=0.01$. It seems that results for linear system sizes $N=4,8$
and 16 have converged around $T_c/J\cong 1.05$, but as the system size
is increased strong corrections appear and push $T_c/J$ down to around
$T_c/J\cong 0.9$.  This is a clear case where it would be dangerous to
draw conclusions from a study of small system sizes. The results for
system size $N=256$ have statistical errors which are slightly larger
than the symbol size, while all other statistical errors are much
smaller. To determine $T_c$ very accurately for such a small
anisotropy one would need accurate data for even larger system
sizes. However, the current precision is enough for the comparisons
with results of approximate theoretical
 methods that we present in
the next section.

\subsection{Magnetization}

We compare the QMC data with Schwinger Boson mean-field\cite{timm00}
and many-body Green's functions calculations.\cite{frob00,frob02} Both
methods include spin wave excitations approximately and represent
significant improvements over simple mean field theories in which
magnon excitations are neglected completely.  Of these two methods the
Schwinger boson theories are numerically less demanding and much
easier to extend to arbitrary spin than the RPA approach.\cite{timm00}

In the Schwinger boson theory the Heisenberg model is mapped onto an
equivalent bosonic system. This can be done by using the SU(2)
symmetry in spin space of the Heisenberg model. The SU(2) model can
then be generalized to an SU($N$) model, containing $N$ bosons per
site. In the limit $N\rightarrow\infty$ mean-field theory becomes
exact, and in this section we label mean-field
SU($N\rightarrow\infty$) results by ``SU($N$)''. Using the local
equivalence between the SU(2) and O(3) groups, this can be repeated
for an O(3) model, and we label mean-field O($N\rightarrow\infty$)
results by ``O($N$)''.

\begin{figure}
\resizebox{!}{6cm}{\includegraphics{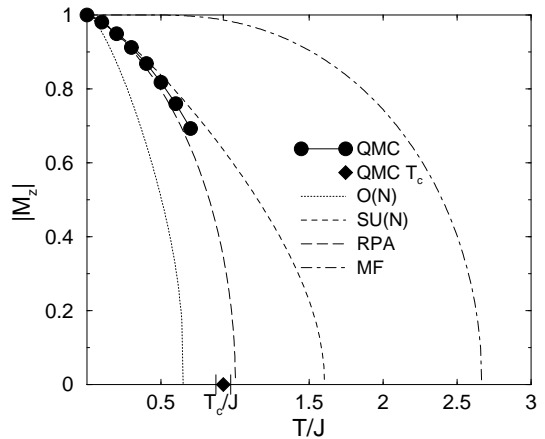}}
\caption{\label{fig:mag001} Magnetization as a function of temperature
$T/J$ for $S=1$ and $K_2/J=0.01$.}
\end{figure}

\begin{figure}
\resizebox{!}{6cm}{\includegraphics{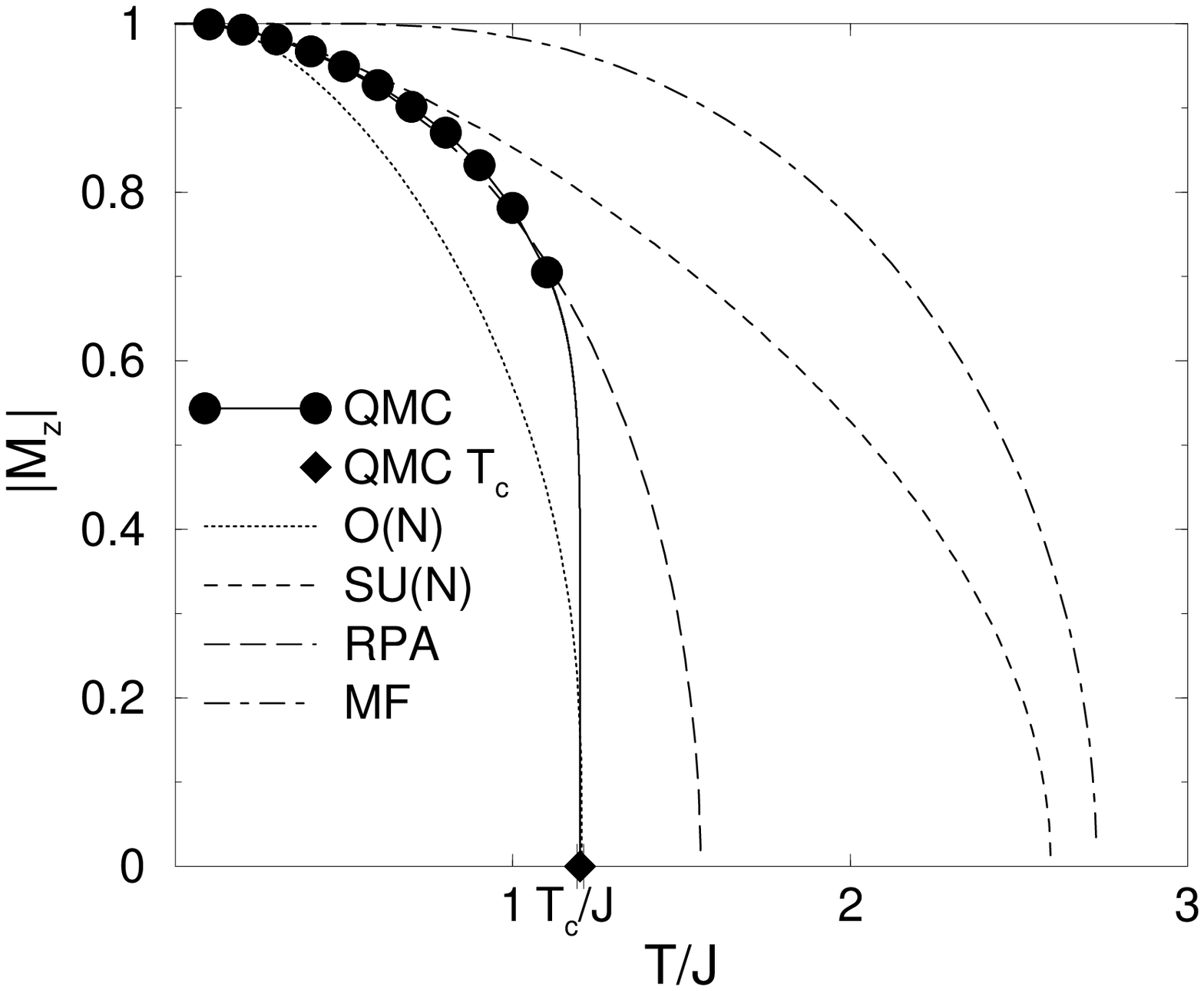}}
\caption{\label{fig:mag020} Magnetization as a function of temperature
$T/J$ for $S=1$ and $K_2/J=0.20$.}
\end{figure}

\begin{figure}
\resizebox{!}{6cm}{\includegraphics{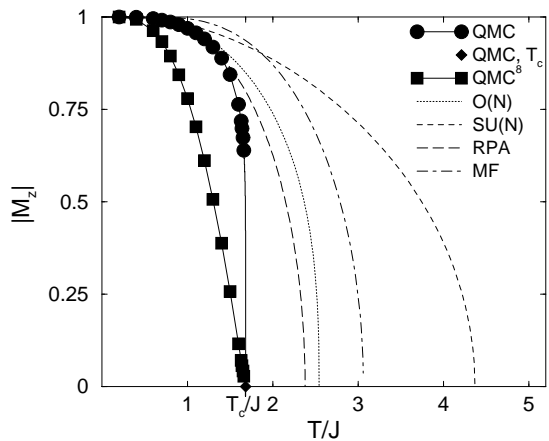}}
\caption{\label{fig:mag150} Magnetization as a function of temperature
$T/J$ for $S=1$ and $K_2/J=1.50$.}
\end{figure}

The many-body Green's function calculations for the magnetization are
done by a procedure where one works at the second level of the
hierarchy of equations of motion for the Green's
functions.\cite{frob02} This allows an exact treatment of the terms
stemming from the single-ion anisotropy, whereas the exchange
interaction terms are treated by the Tyablicov\cite{tyab59}, or
random-phase approximation (RPA), and from now on Green's function
results will be labeled by RPA. This procedure is an improvement over
the Anderson-Callen decoupling\cite{frob00}, in which the single-ion
anisotropy terms are decoupled at the level of the lowest-order
equation of motion, which is a good approximation only for small
anisotropies.\cite{frob02} The results in this subsection are
calculated using the exact treatment of the single-ion
anisotropy. Owing to problems with numerical stability, the
calculations for the reorientation of the magnetization discussed in
the next subsection are done with the Anderson-Callen decoupling.

We compare magnetization curves for three values of the anisotropy
covering three orders of magnitude, $K_2/J=0.01$, 0.2 and 1.5. The
smallest value is of greatest relevance for experiments, but it is of
interest to see how well the analytic methods work outside this region
as well.  In Fig.~\ref{fig:mag001} we see that, for $K_2/J=0.01$ the
RPA calculations give a rather accurate magnetization curve, while the
O($N$) and SU($N$) theories give a low and high estimate,
respectively. In the low-temperature limit both the SU($N$) theory and
the RPA calculation recover the correct spin-wave result. Note also by
how much a simple mean-field (MF) theory overestimates the
magnetization ($T_c^{\mathrm{MF}}\simeq 2.7\ T_c^{\mathrm{QMC}}\
!$). As is well known, mean-field theory also totally fails at low
temperatures (exponential instead of power law behaviour), due to the
neglect of spin waves.  In Fig.~\ref{fig:mag020} we see that for
$K_2/J=0.2$ the RPA still gives very accurate values at low
temperatures, while overestimating the magnetization at higher
temperatures. The O($N$) theory happens to give a good estimate of
$T_c$, while the SU($N$) again overestimates the magnetization. In
Fig.~\ref{fig:mag020} the QMC points are connected by straight lines,
except for the line between $T_c$ and the highest temperature below
$T_c$. This line is a fit to the Ising-like critical behavior
\begin{equation}
M_z\propto (T-T_c)^{\frac{1}{8}}.
\end{equation}
The good fit indicates that the QMC data have come close enough to
$T_c$ for the critical behavior to set in.  For the largest
anisotropy, $K_2/J=1.5$ (Fig.~\ref{fig:mag150}), the RPA calculation
again gives the most accurate estimate, but all approximate curves
result in too high a magnetization. For high temperatures the SU($N$)
theory yields a larger magnetization than simple mean-field theory.
In Fig.~\ref{fig:mag150} we have also included the quantum Monte Carlo
data raised to the eighth power. The resulting straight line is
further evidence of how close to the critical point the simulation has
come.

The large deviations of the approximate theoretical treatments from
QMC, in particular close to $T_c$ and for large anisotropies, indicate
that neither the Green's function approach nor the Schwinger boson
method treats spin wave interactions in a satisfactory way.  It is
interesting to note that that RPA and SU($N$) give a (mean-field)
exponent of $1/2$, whereas O($N$) gives $1/3$.

In this subsection we have compared QMC data with approximate
theoretical results for the magnetization curve. Except for the value
of $T_c$, the QMC data shown above have converged in system size
within error bars that are not discernible in the figures. The largest
system size used is $256\times 256$ spins. Error bars are shown for
the Binder estimate of $T_c$. As can be seen the error increases with
decreasing anisotropy, which is to be expected.

\subsection{Reorientation}

We consider two examples for the reorientation of the magnetization in
a transverse field. For an anisotropic spin-2 model we have calculated
the vertical and transverse components of the magnetization, $M_z$ and
$M_x$, as a function of a transverse magnetic field at
 a fixed
temperature.

The anisotropy favors an out-of-plane magnetization, while the
transverse field wants the magnetization to be in the plane. This
competition results in quite interesting phase diagrams, where the
order of the transition in general depends on the order of the
anisotropy.\cite{timm00} Here we concentrate on two examples, to
demonstrate the applicability of the QMC approach.

\begin{figure}
\resizebox{7cm}{7cm}{\includegraphics{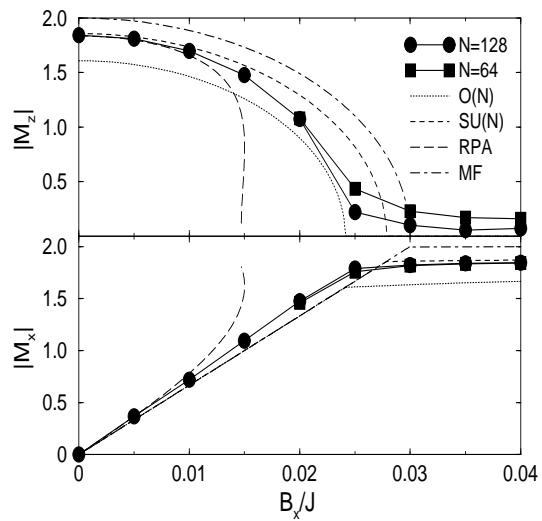}}
\caption{\label{fig:field001} In-plane and out-of-plane magnetization as
a function of transverse field at temperature $T/J=1 $ for $S=2$,
$K_2/J=0.01$. }
\end{figure}

\begin{figure}
\resizebox{7cm}{7cm}{\includegraphics{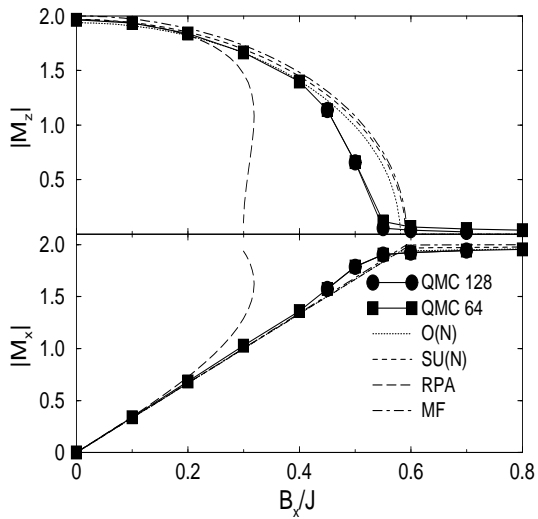}}
\caption{\label{fig:field020} In-plane and out-of-plane magnetization as
a function of transverse field at temperature $T/J=1 $ for $S=2$,
$K_2/J=0.2$. }
\end{figure}

In Fig.~\ref{fig:field001} we show the magnetization curves for a
small anisotropy, $K_2/J=0.01$, at a temperature $T/J=1$.  Primarily
we show QMC results for system size $128\times 128$, but in addition
we show results for $64\times 64$ in the region where they differ from
each other. For small and intermediate fields the results have
converged, while we again see increasing finite-size effects closer to
the critical field. The trend is, however, clear. The transverse
magnetization does appear to increase linearly with the transverse
field, up to the highest fields for which the results have converged.

The SU($N$) and O($N$) calculations both show a linear dependence on
the transverse magnetization, $M_x$, similar to QMC, whereas the
vertical component is over- or underestimated, respectively.  The
SU($N$) results for $M_z$ are good at weak field, though. The
mean-field results for $M_x$ are also linear but mean-field theory
consistently overestimates $M_z$. The RPA calculation with the
approximate decoupling of the anisotropy terms\cite{frob00} follows
the QMC curves at small values of the transverse field, but the
reorientation occurs at a considerably smaller critical field than in
QMC and in the other approximations. Also, in the RPA the
magnetization is not a unique function of field close to the
transition. The same system has been studied for a larger anisotropy,
$K_2/J=0.20$, in Fig.~\ref{fig:field020}. Here the SU($N$) and O($N$)
solutions are mean-field-like and agree very well with the QMC
solution, except close to the transition, whereas the RPA behaves as
in Fig.~\ref{fig:field001}. Close to the transition the SU($N$),
O($N$), and mean-field solutions overestimate the vertical
magnetization. The QMC results display a small deviation from the
linear increase in the in-plane magnetization, which is not reproduced
by the Schwinger boson or mean-field methods.

 \section{\label{sec:con}Conclusion}

This work shows the feasibility of using large-scale QMC calculations
to examine microscopic thin film models. The QMC approach can be used
both as a sanity check on approximate theoretical treatments and as a
method in its own right.  Our results indicate that in the absence of
a transverse field, RPA with an exact treatment of the anisotropy
terms\cite{frob02} appears to be more accurate than the Schwinger
boson calculation. Both methods, however, represent an improvement
over simple mean-field theory for small anisotropies but QMC reveals
weaknesses in the approximate theories at large anisotropies and close
to the Curie temperature. In the presence of a transverse field the
results obtained from Schwinger boson methods are quite
mean-field-like, which turns out to be appropriate for most of the
field range. The reason that the RPA is worse in this case is probably
due to the Anderson-Callen decoupling of the anisotropy terms in
RPA.\cite{frob00} This was necessary because the more accurate
treatment\cite{frob02} led to numerical difficulties when applied to
the reorientation problem. To make more definite conclusions about the
merits of different approximate methods, we would have to investigate
a much larger parameter space, including higher anisotropies, general
spin, temperature, and transverse field and the extension to several
layers. We leave such an investigation as work for the future.

It is possible to extend the QMC calculations to include several
layers with arbitrary inter-layer coupling, as well as, for example,
anisotropies in the exchange coupling. Unfortunately, the dipole
interaction introduces frustration and thereby the sign problem, and
is therefore currently out of reach for QMC studies. On the other
hand, long-range ferromagnetic couplings are not a problem.  The
implementation of the dipole coupling also leads to problems in the
Schwinger boson theory but is possible in a Green's function
description.\cite{jens99,frob00} A very exciting recent development in
QMC is the introduction of directed loop moves,\cite{sylj02} which,
according to our initial calculations, can reduce the autocorrelation
time by one order of magnitude in spin-1 systems, and therefore make
it possible to reach substantially larger system sizes. However, it is
not clear that the directed loops are as easy to implement for a
general model as the method used in this work.

We are grateful to A. Sandvik for fruitful
discussions. P.H. acknowledges support by the Swedish Research Council
and Ella och Georg Ehrnrooths stiftelse.

\bibliography{bib}

\end{document}